\documentstyle[prl,aps,epsfig]{revtex}
\def\ket#1{\vert#1\rangle}
\def\bra#1{\langle#1\vert}
\def\brak#1#2{\langle#1\vert #2\rangle}

\def\bp{{\bf{p}}}
\def\bk{{\bf{k}}}
\def\bpi{{\bf{\pi}}}
\def\bxi{{\bf{\xi}}}

\def\apd{\hat a^{\dag} (\bp)}

\def\be{\begin{equation}}
\def\ee{\end{equation}}
\def\bea{\begin{eqnarray}}
\def\eea{\end{eqnarray}}
\def\dst{\displaystyle\phantom{|}}
\def\ov{\over\dst}
\def\bak{{\bf K}}
\def\dek{{\bf \Delta k}}
 
\begin{document}
\draft
\title{Analytic Solution of the Pion-Laser Model}
\author{T. Cs\"org\H o$^{1,2}$
 and J. Zim\'anyi$^{1}$
} 
\address{
{$^1$Department of Physics, Columbia University,
 538 W 120-th Street, New York, NY 10027 }\\
{$^2$MTA KFKI RMKI, H-1525 Budapest 114. POB. 49, Hungary}
 }
\date{\today}
\maketitle
\begin{abstract}
  Brooding over bosons, wave packets and Bose - Einstein
correlations, we find that a generalization of the pion-laser model 
for the case of overlapping wave-packets is analytically
solvable with complete $n$-particle symmetrization.
Multi-boson correlations generate  {\it momentum-dependent}
radius and intercept parameters even for {\it static} sources.
Explicit multiplicity dependence
of exclusive correlations and spectra is found. 
The HBT radii are  {\it reduced  for low values }
and {\it enlargened for high values of the mean momentum}
in the rare gas limiting case. 
\end{abstract}
\pacs{25.75.gz,25.75.-q,03.65.-w,05.30.Jp}
\underline{\it Introduction.}
The study of the statistical properties of quantum systems has a long
history with important recent developments. In high energy physics,
quantum statistical correlations are studied in order to infer the
space-time dimensions of the elementary particle reactions.
In high energy heavy
ion collisions hundreds of bosons are created in the present CERN SPS
reactions when $Pb + Pb$ reactions are measured at 160 AGeV laboratory
bombarding energy. At the RHIC accelerator, to be completed 
by 1999, thousands of pions
could be produced in a unit rapidity interval~\cite{qm96,s96}. 
If the number of pions 
in a unit value of phase-space is large enough these bosons may condense
into the same quantum state and a pion laser could be created~\cite{plaser}.
Similarly to this process, when a large number of bosonic atoms
are collected in a magnetic trap and cooled down to increase their
density in phase-space, the bosonic nature of the atoms reveals itself
in the formation of a Bose-Einstein condensate~\cite{atom},
a macroscopic quantum state. Such a condensation mechanism
may provide the key to the formation of atomic lasers in condensed matter
physics and to the formation of pion lasers in high energy particle and
heavy ion physics, reviewed recently in refs~\cite{bengt,bill,uli_sum}.

The density matrix of a generic quantum mechanical 
system is 
\be
\hat \rho = \sum_{n=0}^{\infty} \, {p}_n \, \hat \rho_n,
\ee
Here the index $n$ characterizes sub-systems with 
particle number fixed to $n$, 
the multiplicity distribution is prescribed by
the set of $\left\{ {p}_n\right\}_{n=0}^{\infty}$, normalized as
$\sum_{n= 0}^{\infty} p_n = 1$. 
The density matrixes are normalized as 
$\mbox{\bf Tr} \, \hat \rho = 1$ and $\mbox{\bf Tr} \, \hat \rho_n = 1$,
where
\be
\hat \rho_n \!\! = \!\! \int \!\! d\alpha_1 ... d\alpha_n
\,\,\rho_n(\alpha_1,...,\alpha_n)
\,\ket{\alpha_1,...,\alpha_n} \bra{\alpha_1,...,\alpha_n}
\ee
and the states $\ket{\alpha_1,...,\alpha_n}$ denote properly normalized
$n$-particle wave-packet boson states. 
 
A wave packet creation operator is
\bea
\alpha_i^{\dag} \! & = &\int\!\! {d^3\bp \over
(\pi \sigma^2 )^{3\over 4} } \! 
	\mbox{\rm e}^{\! -{( \bp- \bpi_i)^{2}\over 2 \sigma_i^{2}}
	-  i \bxi_i (\bp - \bpi_i) + 
 	 i \omega(\bp) ( t-t_i) } \ \apd ,
\label{e:4}
\eea 
where $\alpha_i = (\bxi_i, \bpi_i, \sigma_i,t_i)$ refers 
to the parameters of the
wave packet $i$: the center in space, in momentum space, the width
in momentum space and the production time, respectively. 
For simplicity we assume that all the wave packets are 
emitted at the same instant and with the same width,
$\alpha_i = (\bpi_i, \bxi_i, \sigma, t_0)$.

The $n$ boson states, normalized to unity, are given as
\be
\ket{\ \alpha_1, \ ...\ , \ \alpha_n} = 
 {\left({ \displaystyle{\strut \sum_{\sigma^{(n)} }
	\prod_{i=1}^n \brak{\alpha_i}{\alpha_{\sigma_i}} } } 
	\right)^{- {1\over 2}} } \!\!
\  \alpha^{\dag}_n  \ ... \
\alpha_1^{\dag} \ket{0}.
\label{e:expec2}
\ee
Here $\sigma^{(n)}$ denotes the set of all the permutations of 
the indexes $\left\{1, 2, ..., n\right\}$ 
and the subscript sized $_{\sigma_i}$ denotes the index that
replaces the index $_i$ in a given permutation from $\sigma^{(n)}$.

\underline{\it Solution for a New Type of Density Matrix \label{s:3}}
There is one special density matrix, for which one can
overcome the difficulty, related to the
non-vanishing overlap of many hundreds of wave-packets,
even in an explicit analytical manner. 
Namely, if one assumes, that we have
a system, in which the emission probability  of a boson is increased if
there is an other emission in the vicinity:
\be
\rho_n(\alpha_1,...,\alpha_n) \! = \! 
	{\dst 1 \ov {\cal N}{(n)}} \!
 \prod_{i=1}^n \rho_1(\alpha_i)  \!
\left(\sum_{\sigma^{(n)}} \prod_{k=1}^n \, 
\brak{\alpha_k}{\alpha_{\sigma_k}}
\right)
\label{e:dtrick}
\ee
The coefficient of proportionality, ${\cal N}{(n)}$,
 can be determined from the normalization condition.
 The density matrix of eq.~(\ref{e:dtrick}) describes a 
 quantum-mechanical wave-packet system with induced emission, and the
 amount of the induced emission is controlled by the overlap of the $n$
 wave-packets~\cite{jzcst}, yielding a weight in the range of $[1,n!]$.
 Although it is very difficult numerically to operate with such a 
 wildly fluctuating weight, we were able to reduce the problem~\cite{jzcst}
 to an already discovered ``ring" - algebra of permanents for plane-wave
 outgoing states~\cite{plaser}.
 
For the sake of simplicity we assume 
a non-relativistic, non-expanding
static source at rest in the frame where the calculations are performed. 
\bea
\rho_1(\alpha)& = &\rho_x(\bxi)\, \rho_p (\bpi)\, \delta(t-t_0), \cr
\rho_x(\bxi) & = &{1 \over (2 \pi R^2)^{3\over 2} }\, \exp(-\bxi^2/(2
R^2) ), \cr
\rho_p(\bpi) & = &{1 \over (2 \pi m T)^{3\over 2} }\, \exp(-\bpi^2/(2 m
T) ),
\eea
and a Poisson multiplicity distribution $p_n^{(0)}$ for the case when
the Bose-Einstein effects are negligible:
\be
        {p}^{(0)}_n = {n_0 ^n \over n!} \exp(-n_0),
\ee
 This corresponds to the very rare gas limit,
and completes the specification of the model.
The plane-wave model, to which the multi-particle
wave-packet model was reduced in ref.~\cite{jzcst},
 can be further simplified~\cite{jzcst} to a set of
recurrence relations with the help 
of the so-called ``ring-algebra" discovered first
by S. Pratt in ref.~\cite{plaser}. 
The probability of finding events with multiplicity $n$,
as well as the single-particle and the two-particle momentum distribution in
such  events is given as 
\bea
	{p}_n  & = &
		\omega_{n} \left( \sum_{k=0}^{\infty} \omega_{k} \right)^{-1},
			\label{e:d.1}\\
	N^{(n)}_1(\bk_1)  &  = & 
                \sum_{i=1}^n  {\dst  \omega_{n-i} \ov \omega_{n}}
		 G_i (1,1) , \label{e:d.2} \\
	N^{(n)}_2(\bk_1,\bk_2)   &  =  & 
		\sum_{l=2}^n 
		\sum_{m=1}^{l-1} 
		{\dst \omega_{n-l}  \ov \omega_n}
		\left[ G_m(1,1) G_{l-m} (2,2) +
		 \right.\nonumber \\
	\null &  \null & \qquad\qquad \left.
		+ G_m(1,2) G_{l-m}(2,1)\right] 
		, \label{e:d.3}
\eea
where $\omega_n = p_n / p_0$ and
\be
        G_n(i,j)  =  
		n_0^n \, h_n \exp(-a_n (\bk^2_i + \bk^2_j) + g_n \bk_i \bk_j).
                        \label{e:3.7}
			\label{e:a.1}
\ee
Averaging over the multiplicity distribution $p_n$ yields the inclusive
spectra as
\bea
	G(1,2) & = & \sum_{n=1}^{\infty} G_n(1,2), \\
	N_1(\bk_1) & = & \sum_{n=1}^{\infty} p_n N^{(n)}_1(\bk_1)
		\, = \,  G(1,1), \\
	N_2(\bk_1,\bk_2) & = & G(1,1) G(2,2) + G(1,2) G(2,1).
\eea
An auxiliary quantity is introduced as
\be 
        C_n \!  = \! {\dst 1 \ov n}\! \int \! d^3 \bk_1 
		\, G_n(1,1) \label{e:3.4} 
             	\, = \, h_n \! {\dst n_0^n \ov n }\!
		\left( {\dst \pi \ov{2 a_n -g_n}}\right)^{3\over 2}. 
\ee
With the help of the notation
\bea
        \sigma_T^2  =  \sigma^2 + 2 m T, 
	\label{e:sigt}
	& \qquad & 
        R_{e}^2  =  R^2 + {\dst m T \ov \sigma^2 \sigma_T^2},  \label{e:reff}
\eea
the recurrence relations that
correspond to the solution
of the ring-algebra~\cite{chao,plaser} 
are obtained~\cite{jzcst}
for the case of the multi-particle wave-packet model. These 
correspond to the pion laser model of 
S. Pratt when a replacement $ R \rightarrow R_{e}$ 
and $T \rightarrow T_{e} = \sigma_T^2/(2 m)$ is performed. 

Let us introduce the following auxiliary quantities:
 \bea
	\gamma_{\pm} 
		 =  {\dst 1\ov 2} \left( 1 + x \pm \sqrt{1 + 2 x} \right)
			\label{e:gam.s} & \qquad  & 
	x  =  R_{e}^2 \sigma_T^2
			\label{e:di.x}
 \eea
	The {\it general analytical solution} of the model is given through
	the generating function of the multiplicity distribution $p_n$ 
 \bea
	G(z) & = & \sum_{n = 0}^{\infty} p_n z^n  \label{e:d.g}
	\, = \, \exp\left( \sum_{n=1}^{\infty} C_n (z^n - 1) \right),
	\label{e:gsolu}
 \eea
	where $C_n$ is 
 \be
	 C_n  =  {\dst n_0^n \ov n} 
			\left[ \gamma_+^{n\over 2} - 
				\gamma_-^{n\over 2} \right]^{-3 },
		\label{e:c.s} 
 \ee
	together with the {\it general analytic solution} 
	for the functions $G_n(1,2)$:
 \bea
	G_n(1,2)\! & = & \! j_n 
	\mbox{\rm e}^{ 
	- {b_n \over 2} \left[ 
		\left(\gamma_+^{n\over 2} \bk_1 
		- \gamma_-^{n \over 2} \bk_2\right)^2
		+ \left(\gamma_+^{n\over 2} \bk_2 - 
		  \gamma_-^{n \over 2} \bk_1\right)^2 \right]
	},\!\! \\
	j_n \! & = & \! n_0^n\! \left[{ b_n \over \pi}\right]^{3 \over 2} 
	\qquad 
	b_n \,  = \, {\dst 1 \over  \sigma_T^2} 
			{\dst \gamma_+ - \gamma_- \ov \gamma_+^n - \gamma_-^n} 
 \eea
The detailed proof that 
 the analytic solution to the  multi-particle wave-packet  model is indeed
given by the above equations is described in ref.~\cite{jzcst}.

	The representation of eq.~(\ref{e:gsolu}) indicates that 
	the quantities $C_n$-s are the so called combinants
	~\cite{gyul-comb,hegyi-c1,hegyi-c2} of the
	probability distribution of $p_n$ and in our case their
	explicit form is known for any set of model parameters, as 
	given by eqs. 
	~(\ref{e:c.s},\ref{e:di.x},\ref{e:reff}). 
	The resulting multiplicity generating function 
	does not correspond to discrete probability generating functions
	in ref~\cite{johnson}, we have found a new type of
	probability generating functions. 

	One can prove~\cite{jzcst}, that the mean multiplicity
	 $ \langle n \rangle = 
	\sum_{n=1}^{\infty} n p_n =  \sum_{i = 1}^{\infty} i C_i$.
	The large $n$ behavior of $n C_n $ depends on the ratio of
	$ n_0 / \gamma_+^{3\over 2}$, since for large values of $n$,
	we always have $ (\gamma_- / \gamma_+)^{n\over 2} << 1$. 
	The critical value of $n_0$ is
 \bea
	n_c & = & \gamma_+^{3\over 2} = \left[ {\dst 1 + x + \sqrt{1 + 2 x} 
			\ov 2 } \right]^{3\over 2}.
 \eea
	If $n_0 < n_c$, 
	one finds $lim_{n \rightarrow \infty} n C_n = 0$
	and $\langle n \rangle < \infty$, 
	if  $n_0 >  n_c$ 
	one obtains $lim_{n \rightarrow \infty} n C_n = \infty $ 
	and $\langle n \rangle = \infty$, 
	finally,
	if $n_0 = n_c$ one finds
	$lim_{n \rightarrow \infty} n C_n = 1$ 
	and $\langle n \rangle = \infty$. 
	The divergence of the mean multiplicity
	$ \langle n \rangle $ is related to condensation of the 
	wave-packets to the wave-packet state with $\bpi = 0$,
	i.e. to the wave-packet state with zero mean momentum~\cite{jzcst},
	if $n_0 \ge n_c$.

	The multiplicity distribution of eq.~(\ref{e:gsolu}) is
	studied at greater length in ref.~\cite{jzcst}.

 \underline{\it Rare gas limiting case}.
 Large source sizes or 
 large effective temperatures correspond to
 the $ x >> 1 $ limiting case,
 where the general analytical solution of the model, presented above,
 becomes particularly simple and the exclusive and inclusive spectra
 and correlation functions can be obtained analytically 
	to leading order in $ 1/x << 1$.
 From eq.~(\ref{e:a.1}) one obtains that
 \bea
	G_n(1,2) & = &  j_n 
		\exp\!\!\left[ 
		 - {\dst n \ov 2 \sigma_T^2 }( \bk_1^2 + \bk_2^2) 
		- \! {R_{e}^2 \ov 2 n} 
			{\bf \Delta}\bk^2 \right]\!, \\
	\label{e:g.i.rare} 
	j_n & = &{\dst n^{5/2} C_n \ov (\pi \sigma_T^2)^{3\over 2} },
	\qquad 
	C_n \,\, = \,\, { \dst n_0^n  \ov n^4 } 
		\left({\dst 2 \ov x} \right)^{{3\over 2}(n-1)}\!\!,
	\label{e:c.rare}
 \eea
	where 	${\bf \Delta}\bk = \bk_1 - \bk_2$.
	We can see from eq.~(\ref{e:g.i.rare}), that
	the higher order corrections will contribute to the observables
	with reduced effective temperatures and reduced effective radii.
	Eq. ~(\ref{e:c.rare}) indicates, 
	that the leading order result for the combinants 
	in the $ x >> 1 $ limiting case is $C_1$ with the first 
	sub-leading correction given by $C_2$. 
	Thus, the probability distribution can be considered
	in the rare gas limiting case
 	as a Poisson distribution of particle
	 singlets with a sub-leading correction
 	that yields a convolution of Poisson-distributed doublets.

	The {\it very} rare gas limiting case  corresponds to keeping only the
	leading $n=1$ order terms in the above equations.
	The multiplicity distribution
	is a Poisson distribution with
	$\langle n \rangle = n_0$ and no influence from stimulated emission.
	The momentum distribution is a static Boltzmann
	distribution, and the exclusive and inclusive momentum distributions
	coincide~\cite{jzcst}. The leading order two-particle Bose-Einstein
	correlation function is a static Gaussian correlation function
	with a constant intercept parameter of $\lambda = 1$ and
	with a momentum - independent radius parameter of $R_* = R_{e}$
	~\cite{jzcst}.

	This wave-packet 
	model exhibits a lasing behavior in the very dense
	Bose-gas limit, which corresponds to an optically coherent
	behavior, characterized by a vanishing enhancement of the
	two-particle intensity correlations at low momentum,
	$C(\bk_1,\bk_2) = 1$, a case which 
	is described in greater details in ref.~\cite{jzcst}.

	The probability generating function yields 
	the following leading order multiplicity distribution:
\be
	p_n = {\dst n_0^n \ov n!} \exp(-n_0) \,
		\left[ 1 + {\dst n(n-1) - n_0^2 \ov 2 (2 x)^{3\over 2}
			} \right].
	\label{e:pn.sol}
\ee

	The mean multiplicity, the factorial cumulant moments of
	the multiplicity distribution, the inclusive and exclusive
	momentum distributions were evaluated by keeping only the
	leading order  terms in $1/x$ in
	ref.~\cite{jzcst}. 
	The two-particle exclusive
correlation functions can also be evaluated by 
applying a Gaussian approximation to the
leading order corrections in the $x >> 1$ limiting case:
\bea
C^{(n)}_2(\bk_1,\bk_2) \!\! & = &   {\dst n^2 \ov n(n-1) } 
		{\dst N_2^{(n)}(\bk_1,\bk_2) \ov
		N_1^{(n)}(\bk_1) \, N_1^{(n)}(\bk_2) }
		  \nonumber \\
\null & = & \!\! 1 + \lambda_{\bak} \!
		\exp\!\left( - R_{\bak,s}^2  \dek_{s}^2 
		- R_{\bak,o}^2 \dek_{o}^2 \right), \!
\eea
where $\bak = 0.5 (\bk_1 + \bk_2)$, the side and outwards directions
are introduced utilizing the spherical symmetry of the source as
$\dek_{s} = \dek - \bak{ (\dek \cdot \bak) / ( \bak \cdot \bak)} $ and 
$\dek_{o} = \bak{ (\dek \cdot \bak) / ( \bak \cdot \bak)} $,
 similarly to refs.~\cite{nr}. The mean-momentum dependent intercept
and radius parameters are 
\bea
\lambda_{\bf{K}} \!\! & = & \! \!  1 +
		{\dst 2 \ov ( 2 x)^{3\over 2} }
		\left[ 1 - 
		2^{(5/2)} \exp\left( - {\dst {\bf K}^2 \ov \sigma_T^2 } \right)
		\right] \label{e:lambda.corr} \\
R_{{\bf{K}},s}^2 \!\! & =   &\!\!  R_{e}^2  +
		{\dst 1 \ov ( 2 x)^{3\over 2} }
	 	\left[  R_{e}^2 - \sqrt{2} 
		\exp\left( - {{\bf K}^2 \ov \sigma_T^2} \right)	
		\right.\times\nonumber \\
\null & \null & \!\! \qquad \times \left.
		\left( 
		 ( n + 2) R_{e}^2 + {\dst 2 \ov \sigma_T^2 }
		\right)
		\right], \label{e:r.corr}\\
R_{{\bf K},o}^2 \!\! & = & \!\! R_{{\bf{K}},s}^2 +
		{\dst n \ov    x^{3\over 2} } {\dst {\bf K}^2  \ov \sigma_T^4 } 
		\exp\left( - {\dst {\bf K}^2 \ov \sigma_T^2} \right) 
		\label{e:c.dir}
\eea

Thus the symmetrization results in a momentum - dependent 
intercept parameter $\lambda_{\bak}$ that starts from a
$\lambda_{\bak = 0} < 1 $ value at low momentum  and {\it increases} 
with increasing momentum. Already in the first paper about the pion laser
model, ref.~\cite{plaser}, a reduction of the exact intercept parameter
was observed and interpreted as the onset of a coherent behavior in the 
low momentum modes. First, a partially coherent system is created,
characterized by $\lambda_{\bak} < 1$, and if the density of pions
is further increased, one finds  a fully developed pion-laser
with $\lambda_{\bak} = 0$, 
see ref.~\cite{jzcst} for analytic considerations.
 
Observe that the radius parameter at low mean momentum 
decreases while the radius parameter at high mean momentum 
increases, as compared to $R_{e}$. 
{\it
The radius parameter of the exclusive correlation function thus becomes 
momentum-dependent even for static sources!}
This  effect is more pronounced for higher values of the 
fixed multiplicity $n$, in contrast to the momentum dependence of 
$\lambda_{\bak}$ that is independent of $n$.

\begin{figure}
\hbox{
\hskip -0.5cm
\epsfig{file=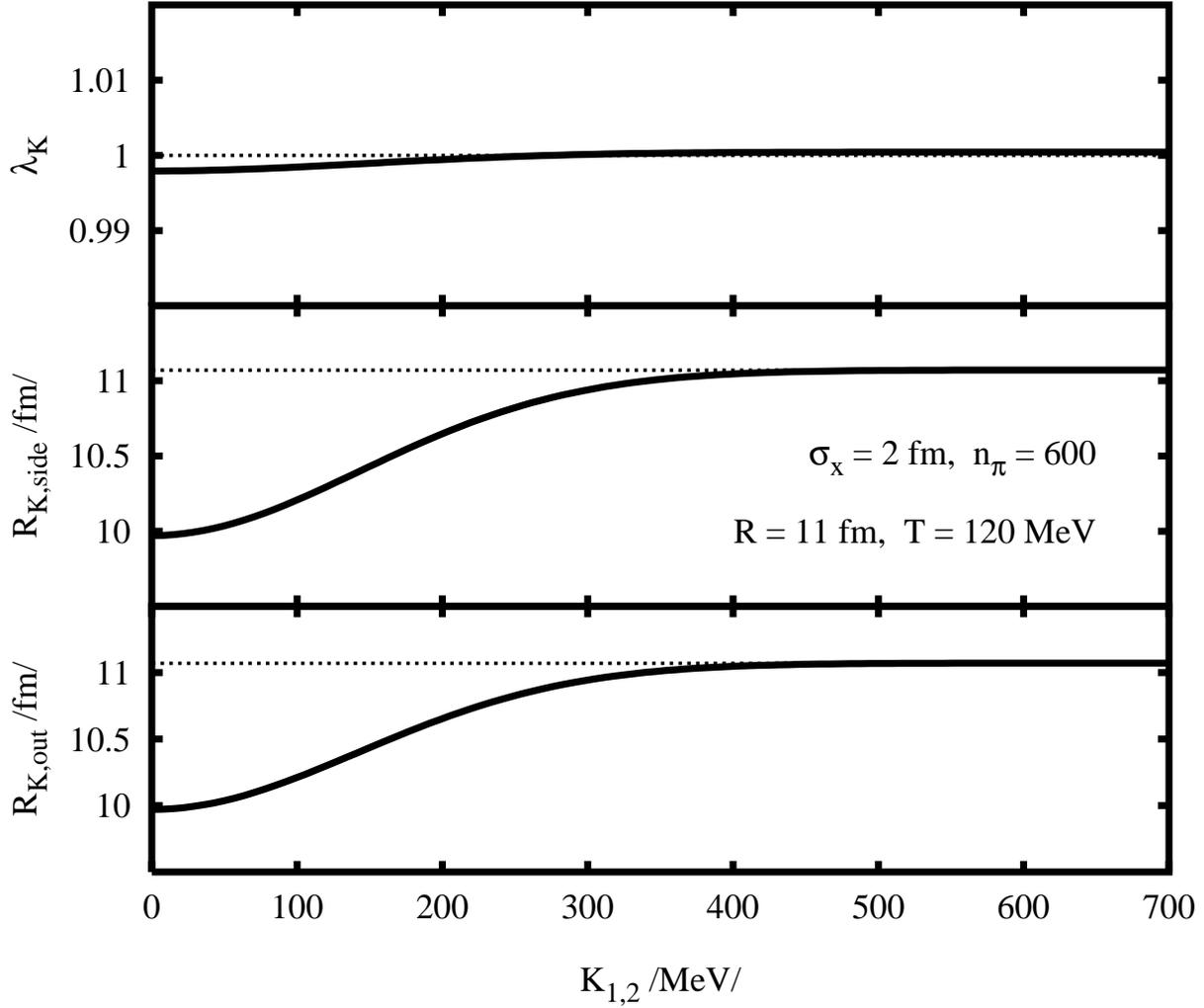,width=17.0cm}
}
\caption{Multi-particle symmetrization results at low $\bak$ in a
	momentum-dependent 
	reduction of the intercept parameter $\lambda_{\bak}$,
	the side-wards and the outwards
	radius parameters, $R_{\bak,s}$ and $R_{\bak,o}$
	from their static values of 1 and $R_{e}$, respectively.
	The enhancement of these parameters at high momentum is
	hardly noticeable for large and hot systems. 
       } 
\label{fig:1}
\end{figure}

Last, but not least, a specific term appears in the two-particle
exclusive correlation function, that contributes only to the {\it
out} direction, which, in case of spherically symmetric sources,
may be identified with the direction of the mean momentum~\cite{nr}.
This directional dependence is related only to the direction of the relative
momentum as compared to the direction of the mean momentum,
and does not violate the assumed spherical symmetry of the boson source.
The effect vanishes both at very low or at very high values of the mean momentum
${\bf K}$, according to eq.~(\ref{e:c.dir}).
The top, middle and bottom panel of Figure ~\ref{fig:1} indicates the 
momentum dependent $\lambda_{\bak}$ intercept parameter,
the $R_{\bak,s}$ and $R_{\bak,o}$ radius parameters for
a fireball with $R = 11 $ fm, $T = 120$ MeV. 
The pions are assumed to be described by wave-packets with
spatial widths of $\sigma_x = 2$ fm, and events with fixed identical pion
multiplicity of $n_{\pi} = 600$ are selected for the evaluation of the
correlation function. For this set of parameters, the enhancement of
$\lambda_{\bak}$, $R_{\bak,side}$ and $R_{\bak,out}$ is hardly
noticeable at high momentum, but their small decrease at low momentum
is clear. One may consider a small cold pionic system with a few large
wave-packets only, by inserting $R = 4 $ fm, $ T = 10 $ MeV,
$n_{\pi} = 3$ and $\sigma_x = 5 $ fm to eqs.
~(\ref{e:lambda.corr}-\ref{e:c.dir}). This source could correspond to
heavy ion collisions in the 30 MeV A energy domain~\cite{nr}
characterized by an 
effective radius $R_{e} = 4.5$ fm and effective temperature of 
$T_{e} = 15.6 $ MeV~\cite{nr}. In this case, the directional
dependence of the radii and the enhancement of the radius parameters
at high momentum as compared to $R_{e}$ becomes significant 
not only analytically but numerically as well.

\underline{\it Highlights :}
In this Letter a consequent quantum mechanical description
of multi-boson systems is presented, using properly normalized
projector operators for overlapping multi-particle wave-packet states
describing stimulated emission of bosons.   
Our new analytic result is 
that multi-boson correlations 
generate  {\it momentum-dependent}
radius and intercept parameters even for {\it static} sources, as well
as induce a special {\it directional dependence} of the correlation
function. The effective {\it radius parameter}  of the two-particle
correlation function {\it is reduced  for low values }
and {\it enlargened for large values of the mean momentum}
in the rare gas limiting case, as compared to the case 
when multi-particle symmetrization effects 
are neglected.  
 For extended, hot and rare gas of a few hundred pions, the
reduction of the radius parameters at low momentum  is found to
be the most apparent effect. The directional dependence of the 
radius parameters and the enhancement of the radii at high momentum
is characteristic for a small, cold pion gas with only a handful of
particles in it.
These results can be understood qualitatively by an enhancement  
of the wave-packets in the low momentum modes,
due to multi-particle Bose-Einstein symmetrization effects,
as the system starts to approach the formation of a laser,
characterized by the appearance of 
partial optical coherence in the low momentum modes.

 Our results explicitly depend on the multiplicity, providing 
{\it a theoretical basis for event-by-event analysis} 
of heavy ion data.
 
\underline{\it Acknowledgments :}
Cs. T. would like to thank M. Gyu\-lassy, S. Hegyi, G. Vahtang  and X. N. Wang 
for stimulating discussions.
This work was supported by the NSF - HAS
Grant INT 8210278, and by the OTKA Grants No.  F4019,
 W01015107 and T024094, by the USA - HJF grant MAKA 378/93 and 
by an Advanced Research Award from the Fulbright
Foundation.  

\end{document}